\documentclass[aps,prb,twocolumn,showpacs,floatfix]{revtex4}

\usepackage{graphicx}
\usepackage{amssymb}

\begin{document}

\title{Simultaneous suppression of ferromagnetism and superconductivity in UCoGe by Si substitution}

\author{D. E. de Nijs}
\author{N. T. Huy}
\author{A. de Visser}
\email{devisser@science.uva.nl}

\affiliation{Van der Waals - Zeeman Institute, University of
Amsterdam, Valckenierstraat~65, 1018 XE Amsterdam, The
Netherlands}

\date{\today}

\begin{abstract}
We investigate the effect of substituting Si for Ge in the
ferromagnetic superconductor UCoGe. Dc-magnetization,
ac-susceptibility and electrical resistivity measurements on
polycrystalline UCoGe$_{1-x}$Si$_x$ samples show that
ferromagnetic order and superconductivity are progressively
depressed with increasing Si content and simultaneously vanish at
a critical concentration $x_{cr} \simeq 0.12$. The non-Fermi
liquid temperature variation in the electrical resistivity near
$x_{cr}$ and the smooth depression of the ordered moment point to
a continuous ferromagnetic quantum phase transition.
Superconductivity is confined to the ferromagnetic phase, which
provides further evidence for magnetically mediated
superconductivity.

\end{abstract}

\pacs{74.70.Tx, 74.62.Dh,75.30.Kz}

\maketitle

Recently, it was discovered~\cite{Huy-PRL-2007} that the
intermetallic compound UCoGe belongs to the small group of
ferromagnetic superconductors (FMSCs): superconductivity with a
transition temperature $T_s = 0.8$ K coexists with weak itinerant
FM order with a Curie temperature $T_C = 3$~K. Ferromagnetic
superconductors attract much interest, because in the standard BCS
scenario superconductivity (SC) and ferromagnetism (FM) are
incompatible~\cite{Berk-PRL-1966}. This is due to the strong
de-pairing effect of the ferromagnetic exchange interaction, which
thwarts phonon mediated formation of singlet Cooper pairs.
However, an alternative route is offered by spin fluctuation
models~\cite{Fay-PRB-1980,Lonzarich-CUP-1997}, in which critical
magnetic fluctuations associated with a ferromagnetic quantum
critical point (FM QCP) mediate SC by pairing the electrons in
triplet states. The FMSCs discovered so far are
UGe$_2$~\cite{Saxena-Nature-2000} (under pressure),
UIr~\cite{Akazawa-JPCM-2004} (under pressure),
URhGe~\cite{Aoki-Nature-2001} and UCoGe~\cite{Huy-PRL-2007}. The
latter two compounds offer the advantage that SC occurs at ambient
pressure, which facilitates the use of a wide range of
experimental techniques to probe magnetically mediated SC.

UCoGe crystallizes~\cite{Lloret-PhDthesis-1988,Canepa-JALCOM-1996}
in the orthorhombic TiNiSi structure (space group $P_{nma}$).
Evidence for the proximity to a FM QCP has been extracted from
magnetization and specific heat measurements~\cite{Huy-PRL-2007}
on polycrystalline samples. The low $T_{C} = 3$ K and the small
value of the ordered moment $m_{0}$= 0.03 $\mu_B$ reveal magnetism
is weak. Itinerant magnetism is corroborated by the small value of
the magnetic entropy~\cite{Huy-PRL-2007} (0.3 \% of $R$ln2)
associated with the magnetic transition. More recently, the
magnetic and SC properties were determined for a
single-crystalline sample~\cite{Huy-arXiv-2008}. Magnetization
data reveal UCoGe is a uniaxial ferromagnet with the ordered
moment $m_{0}$~= 0.07 $\mu_B \simeq 2m_{0}^{poly}$ pointing along
the $c$ axis. The electrical resistivity $\rho (T)$ measured for a
current $I \parallel a$ shows SC below 0.6 K and a sharp kink
signaling the Curie temperature $T_{C} = 2.8~$K. The temperature
variation of the
resistivity~\cite{Huy-arXiv-2008,Huy-tobepublished} is
characteristic~\cite{Moriya-Spinger-1985} for a weak itinerant FM
near a critical point, i.e. a Fermi liquid $\rho \propto T^2$
dependence below $T_{C}$ and scattering at critical FM
fluctuations $\rho \propto T^{5/3}$ there above.

In the generic pressure-temperature phase diagram for FMSCs
~\cite{Fay-PRB-1980,Lonzarich-CUP-1997,Kirkpatrick-PRL-2001,Monthoux-Nature-2007}
the superconducting phase (the dome) is confined to the magnetic
phase and $T_C$ and $T_s$ vanish at the same critical pressure.
Such a phase diagram has been reported for
UGe$_2$~\cite{Saxena-Nature-2000} and UIr~\cite{Akazawa-JPCM-2004}
under pressure. In the case of UCoGe~\cite{Huy-PRL-2007}, the
analysis of the thermal expansion and specific heat data, using
the Ehrenfest relation, shows that $T_C$ decreases with pressure,
whereas $T_s$ increases. This places UCoGe on the far side of the
superconducting dome with respect to the magnetic quantum critical
point. Concurrently, under hydrostatic mechanical pressure $T_s$
is predicted to go through a maximum, before vanishing at the
critical point. In this work we use an alternative route to study
the evolution of FM and SC, namely chemical pressure exerted by
replacing Ge by isoelectronic Si. Ferromagnetic UCoGe and
paramagnetic~\cite{Troc-JMMM-1988} UCoSi are
isostructural~\cite{Lloret-PhDthesis-1988,Canepa-JALCOM-1996}. The
unit cell volume of UCoSi is $\sim 3.5 \%$ smaller than the one of
UCoGe, so chemical pressure is relatively weak. By means of
magnetic and transport measurements we find that FM order and SC
are gradually depressed and vanish simultaneously at a critical
concentration $x_{cr} \simeq 0.12$. SC is confined to the FM phase
in agreement with the generic phase diagram. This yields further
support for magnetically mediated superconductivity.

A series of polycrystalline UCoGe$_{1-x}$Si$_x$ samples were
prepared with $0 \leq x \leq 0.20$ and $x=1$. The constituents
(natural U 3N, Co 4N, Ge 5N and Si 5N) were weighed according to
the nominal composition U$_{1.02}$Co$_{1.02}$Ge$_{1-x}$Si$_x$ and
arc melted together under a high-purity argon atmosphere in a
water-cooled copper crucible. The as-cast samples were annealed
for ten days at 875 $^\circ$C. Samples were cut by spark erosion
in a bar-shape for transport and magnetic measurements. The phase
homogeneity of the annealed samples was investigated by Electron
Probe Micro Analysis (EPMA). The matrix had the 1:1:1 composition
and all samples contained a small amount (2\%) of impurity phases.
The EPMA technique did however not allow for a precise
determination of the Ge and Si ratio, and in the following $x$ is
the nominal concentration. Powder X-ray diffraction patterns at $T
= 300$ K for $x = 0.0, 0.1, 0.2$ and $1.0$ confirmed the TiNiSi
structure. The measured lattice constants are $a = 6.864$~\AA, $b
= 4.196$~\AA ~and $c = 7.261$~\AA~for UCoGe and $a = 6.876$~\AA,
$b = 4.108$~\AA ~and $c = 7.154$~\AA~for UCoSi, in good agreement
with literature~\cite{Lloret-PhDthesis-1988,Canepa-JALCOM-1996}.
The unit cell volume $\Omega$ decreases linearly from
209.5~\AA$^3$ ($x=0$) to 202.1~\AA$^3$ ($x=1$), with the main
contraction along the $b$ and $c$ axis.

The dc-magnetization, $M(T,B)$, was measured in a SQUID
magnetometer in magnetic fields up to 5 T and temperatures down to
2 K. The low-field ($B = 10^{-5}$~T) ac-susceptibility,
$\chi_{ac}$, was measured using a mutual inductance coil and a
phase-sensitive bridge in a $^3$He system with base temperature
0.23 K or in a dilution refrigerator with base temperature 0.02 K.
Electrical resistivity data, $\rho (T)$, were taken using a
low-frequency ac-bridge in a four-point configuration in the same
temperature range.

\begin{figure}
\includegraphics[width=8.5cm]{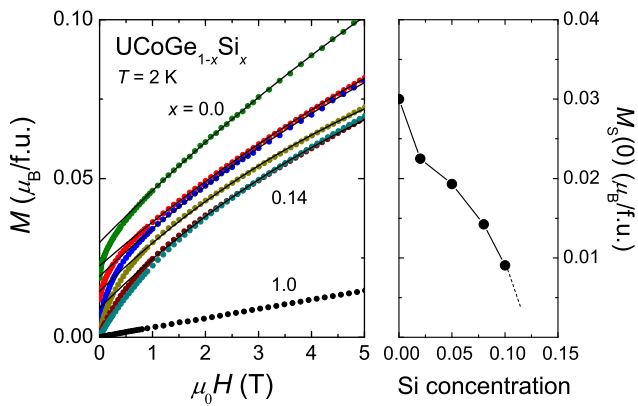}
\caption{(Color online) Left panel: Magnetization as a function of
field for UCoGe$_{1-x}$Si$_x$ alloys at $T = 2$~K. Si
concentrations are (from top to bottom) $x =$ 0, 0.02, 0.05, 0.08,
0.10, 0.14 and 1.0. The solid lines represents fits to Eq.~1 for
$x \leq 0.10$. Right panel: Spontaneous magnetization $M_{s}(0)$
as a function of Si content.}
\end{figure}

The dc-magnetic susceptibility $\chi_{dc} (T)$ of the
UCoGe$_{1-x}$Si$_x$ alloys was measured in an applied field of 1 T
in the temperature range $2-300$ K. The effect of doping small
amounts of Si on $\chi_{dc} (T)$ is weak. For all $x \leq 0.20$
the data for $T = 50-300$~K are described by a modified
Curie-Weiss law, with a temperature independent susceptibility
$\chi_{0} \simeq 10^{-8}$ m$^3$/mol and an effective moment
$p_{eff} \simeq 1.6 \pm 0.1~ \mu_{B}/$f.u.. On the contrary, the
effect of doping on the FM transition is large. Measurements of
the dc-magnetization in a small field ($B=0.01$~T) show that upon
Si doping the FM transition is rapidly suppressed to below the low
temperature limit of our dc-magnetometer (2 K). For $x=$ 0.00 and
0.02 we find $T_C$ = 3.0 K and 2.5 K, respectively. In Fig.~1 we
show the field dependence of the magnetization $M(H)$ measured at
$T=2$~K. The gradual increase of $M(H)$ observed for $B \gtrsim
1$~T is related to the itinerant nature of the magnetic state. The
spontaneous magnetization $M_{s}(H=0)$ rapidly drops with
increasing Si content. For the ordered compounds an estimate of
$M_{s}(0)$ can be made by fitting the data to the empirical
expression
\begin{equation}
M(H) = M_{s}(0) + \Delta M(1-e^{-\mu _0 H/B_0})
\end{equation}
where the parameter $B_0$ probes the magnetic interaction strength
of the fluctuating moments. In the high-field limit $M(H= \infty )
= M_{s}(H=0) + \Delta M$. Eq.~1 describes the experimental data
well for $B \gtrsim 1$~T (solid lines in Fig.~1). The intercepts
of the fits with the vertical axis yield the fit parameters
$M_{s}(H=0)$ in the limit $T \rightarrow 0$. The deviations for $B
< 1$~T are due to the finite temperature at which the data are
taken (the ordered moment is not fully developed yet). For
$x=0.00$~$M_{s}(0) \simeq 0.029 ~\mu_{B}$ ($T \rightarrow 0$) in
agreement with previous results~\cite{Huy-PRL-2007}, while for
$x=0.02 ~M_{s}(0) \simeq 0.022 ~\mu_{B}$. For the samples with
$x=$ 0.05, 0.08 and 0.10 the data have been taken at $T > T_{C}$.
Nevertheless, a rough estimate of $M_{s}(0)$ can be obtained, as
the magnetic transition shows a large temperature broadening in
applied fields $B
> 1$~T. The resulting values of $M_{s}(0)$ are traced in the right
panel of Fig.~1. We conclude $M_{s}(0)$ smoothly goes to zero in
the concentration range $0.10 < x < 0.14$.

\begin{figure}
\includegraphics[width=7cm]{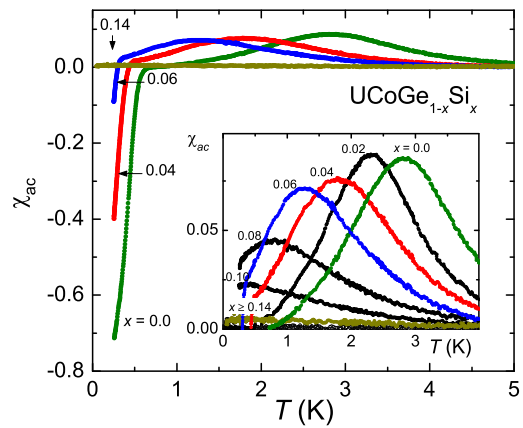}
\caption{(Color online) Temperature variation of the
ac-susceptibility (SI units) of UCoGe$_{1-x}$Si$_x$ alloys for $x
=$ 0.00, 0.04, 0.06 and 0.14. The inset shows $\chi _{ac}$ around
the ferromagnetic transition for $0.00 \leq x \leq 0.20$. The data
for $x=0.20$ fall on the horizontal axis.}
\end{figure}

The suppression of $T_C$ was studied in more detail by the
ac-susceptibility technique. The data, taken down to 0.23 K ($0.00
\leq x \leq 0.10$) and down to 0.02 K ($0.14 \leq x \leq 0.20$),
are shown in Fig.2. The maximum in $\chi _{ac}$ locates the Curie
temperature, which equals 2.8~K and 2.3~K, for $x = $ 0.00 and
0.02 respectively. These values compare well with those extracted
from the dc-magnetization. With increasing Si content the
transition becomes weaker, broadens (see inset in Fig.~2) and for
$x \geq 0.14$ a maximum in $\chi _{ac}$ no longer can be
identified. This confirms magnetism vanishes in the concentration
range $0.10 < x < 0.14$. The large diamagnetic signal measured for
$x =$ 0.00~\cite{Huy-PRL-2007}, 0.04 and 0.06 down to 0.23~K
signals bulk SC. SC is progressively depressed and is no longer
observed for $x=0.14$ (at least down to 0.02 K).

The electrical resistivity was measured in the temperature
interval $0.23-10$ K for $x \leq 0.08$ and in the range $0.02-10$
K for $0.10 \leq x \leq 0.20$. For $x=0.00$ the residual
resistivity $\rho_{0} = 26$~$\mu \Omega$cm. Upon alloying
$\rho_{0}$ increases linearly at least up to $x=0.08$ at the fast
rate of 12~$\mu \Omega $cm per at.\% Si. This shows all Si
substitutes for Ge. Concurrently, the residual resistance ratio
$RRR=R(300$K$)/R(1$K$)$, which amounts to 27 for $x=0$, drops to
$\sim 5$ for $x=0.08$. For $x \geq 0.10$, however, the $RRR$
levels off at a value $\sim 4$. The strong doping sensitivity of
$\rho_{0}$ is possibly related to an enhanced site inversion Ge,Si
$\leftrightarrow$ Co. Notice the TiNiSi structure is an ordered
variant of the CeCu$_2$ structure~\cite{Sechovsky-handbook-1998}
(for U$TX$ compounds crystallizing in the latter structure the
transition metal atoms $T$ and group IV atoms $X$ are randomly
distributed over the $8h$ Cu sites).

\begin{figure}
\includegraphics[width=8.5cm]{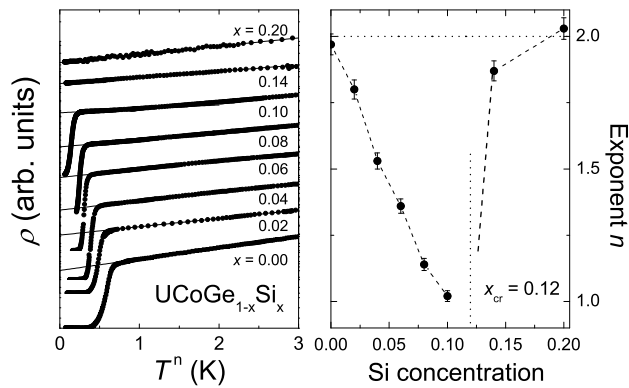}
\caption{Left panel: The electrical resistivity $\rho$ (arbitrary
units) plotted versus $T^n$ of UCoGe$_{1-x}$Si$_x$ alloys for $x$
as indicated. The curves are shifted along the vertical axis for
clarity. The straight solid lines represent fits $\rho \sim T^n$
(see text). Right panel: Exponent $n$ versus Si concentration. The
dashed line serves to guide the eye. The vertical dotted line
locates $x_{cr}$. The horizontal dotted line indicates $n=2$.}
\end{figure}

The FM transition appears as a broad hump in $\rho (T)$ for pure
UCoGe~\cite{Huy-PRL-2007}. Upon alloying, the hump shifts to lower
temperatures at the same rate as the maximum in $\chi_{ac}$. In
Fig.~3 we show the low-temperature part of the resistivity data in
a plot of $\rho$ versus $T^n$. Here $n$ is determined by fitting
$\rho \sim T^n$ for $T_s < T < T_C$. For each $x$ the best value
of $n$ was obtained by fitting over a larger and larger
temperature range, while keeping $n$ constant and the error small.
In the magnetic phase ($x \leq 0.10$) the exponent shows a
quasi-linear decrease from $n=2$ for $x=0.00$ to the
non-Fermi-liquid value $n \simeq 1$ for $x=0.10$ (see Fig.~3).
Close to the critical point the temperature range for the fit
becomes very small and the values of $n$ should be interpreted
with care. Nevertheless, the decreasing trend is evident. For $x
\geq 0.14$ the Fermi liquid value $n=2$ is recovered. The SC
transition is depressed with increasing Si content and no SC has
been observed down to 0.02 K for $x=0.14$.

\begin{figure}
\includegraphics[width=6cm]{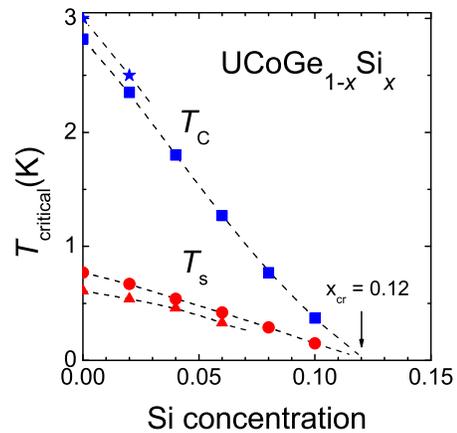}
\caption{(Color online) Curie temperature, determined by $M(T)$
($\bigstar$) and $\chi_{ac}(T)$ ($\blacksquare$), and
superconducting transition temperature, determined by $\rho (T)$
($\bullet$) and $\chi_{ac}(T)$ ($\blacktriangle$), as a function
of $x$ for UCoGe$_{1-x}$Si$_x$ alloys. The dashed lines serve to
guide the eye. Superconductivity and ferromagnetism both vanish at
$x_{cr} \simeq 0.12$.}
\end{figure}

Having determined the evolution of the FM and SC phases in the
UCoGe$_{1-x}$Si$_x$ alloys by magnetic and transport measurements,
we construct the phase diagram shown in Fig.~4. $T_C$ is depressed
quasi-linearly, at least till $x =0.08$, at a rate $dT_{C}/dx =
-0.25$~K/at.\%Si. By extrapolating $T_{C}(x) \rightarrow 0$ we
arrive at a critical Si concentration for the suppression of FM
order $x_{cr} = 0.11$. For $x > 0.08$ a tail appears, and the data
extrapolate to $x_{cr}^{FM} \simeq 0.12$. $T_s$, determined
resistively by the midpoint of the transition, is depressed
somewhat faster than linear, initially at a rate $dT_{s}/dx =
-0.06$~K/at.\%Si. By smoothly extrapolating $T_s (x) \rightarrow
0$ we obtain a critical Si concentration for the suppression of SC
$x_{cr}^{SC} \simeq  0.12$. The $T_s (x)$ values measured by
$\chi_{ac}(T)$ for $x \leq 0.06$, signal the onset of
bulk~\cite{Huy-PRL-2007} SC and follow the same trend. Notice $T_s
(x)$ bulk extrapolates to a slightly lower $x_{cr}$, i.e. close to
the value $x_{cr} = 0.11$ obtained by the linear extrapolation of
$T_{C}(x)$.

In order to compare the effect of chemical and hydrostatic
pressure we calculate from the difference in unit cell volume of
UCoGe and UCoSi that 1~at.\% Si is equivalent to 0.35~kbar (here
we assume the isothermal compressibility $\kappa \simeq
10^{-11}$~Pa$^{-1}$). Concurrently, the measured doping-induced
depression of $T_C$ (Fig.~4) translates to $dT_{C}/dp =
-0.71$~K/kbar, which is about a factor three larger than the value
$-0.25$~K/kbar calculated~\cite{Huy-PRL-2007} via the Ehrenfest
relation. This indicates Si does not merely exert chemical
pressure. Indeed hybridization phenomena in U$TX$ alloys are in
general strongly anisotropic~\cite{Sechovsky-handbook-1998}. As
regards the SC transition, Si doping obviously has a different
effect than hydrostatic pressure. The measured doping-induced
depression of $T_s$ (Fig.~4) translates to $dT_{s}/dp =
-0.17$~K/kbar, while the Ehrenfest relation shows $T_s$ {\it
increases} at a rate $dT_{s}/dp =
0.02$~K/kbar~\cite{Correction_Ts-p}.

The suppression of magnetic order in the UCoGe$_{1-x}$Si$_x$
alloys can be understood in terms of a simple Doniach
picture~\cite{Doniach-PhysicaB-1977}: by doping the smaller Si
atoms the $5f-3d$ hybridization strength increases, which leads to
a loss of magnetism. The rapid suppression of FM order provides
further evidence that UCoGe is close to a FM QCP. This is
corroborated by the steady decrease of the non-Fermi liquid
exponent $n$ of the resistivity measured in the FM phase (see
Fig.~3). The itinerant nature of the FM state suggests that the
critical point is of the
Moriya-Hertz-Millis~\cite{Hertz-PRB-1976,Moriya-Spinger-1985,Millis-PRB-1993}
type. The extracted exponent $n \simeq 1$ near $x_{cr}^{FM}$ is
much smaller than the value $n=5/3$ predicted for a clean FM QCP.
A similar observation was made for the doping-induced FM QCP in
URh$_{1-x}$Ru$_x$Ge alloys~\cite{Huy-PRB-2007}: at $x_{cr}=0.38$
$n \simeq 1.2$. Clearly, disorder
reduces~\cite{Pfleiderer-Nature-2001} $n$. The smooth depression
of $M_{s}(0)$ indicates the ferro-to-paramagnetic transition at $T
= 0~K$ is a continuous phase transition. Additional experiments,
e.g. specific heat, are required to put the evidence for a FM QCP
at $x_{cr} \simeq 0.12$ on firm footing.

The magnetic and SC phase diagram (Fig.~4) presents compelling
evidence that superconductivity is confined to the FM phase.
Moreover, by smoothly extrapolating $T_{C}(x)$ and $T_{s}(x)$ we
arrive at a most important conclusion, namely $x_{cr}^{FM} =
x_{cr}^{SC} \simeq 0.12$. This shows that FM order and SC are
closely tied together. The simultaneous suppression of FM order
and SC yields strong support for triplet SC mediated by FM spin
fluctuations~\cite{Fay-PRB-1980,Lonzarich-CUP-1997,Kirkpatrick-PRL-2001,Monthoux-Nature-2007}.
Evidence for triplet SC is furnished by the absence of Pauli
limiting in the upper critical field
$B_{c2}$~\cite{Huy-arXiv-2008}. Moreover, the observed anisotropy
in $B_{c2}$ provides evidence for an axial SC gap with nodes along
the direction of the ordered moment, as
calculated~\cite{Mineev-PRB-2004} for the $A$ phase of an
orthorhombic FMSC. On the other hand, it is
recognized~\cite{Foulkes-PRB-1977,Fay-PRB-1980} that triplet SC is
extremely sensitive to scattering at non-magnetic impurities and
defects. Therefore, it is surprising that SC survives till doping
concentrations of $\sim$12~at.\%~Si. For our polycrystalline UCoGe
samples, with $RRR \sim 30$, we calculate~\cite{Huy-PRL-2007} an
electron mean free path, $\ell \approx 500$~\AA, in excess of the
SC coherence length $\xi \approx 150$~\AA, a necessary condition
for unconventional SC. Upon replacing Ge by Si the residual
resistance increases, leading to a corresponding decrease of
$\ell$. Unconventional SC therefore would require a strong
doping-induced reduction of $\xi$ as well. The depression of
non-$s$ wave SC by non-magnetic impurities can be modelled using a
generalized form~\cite{Hirschfeld-PRB-1988,Millis-PRB-1988} of the
Abrikosov-Gor'kov pair-breaking theory. A recent example is
provided by the defect-driven depression of $p$-wave SC in the
paramagnet Sr$_2$RuO$_4$~\cite{Mackenzie-PRL-1998}. In the case of
the UCoGe$_{1-x}$Si$_x$ alloys, however, the defect-driven
depression of $T_{s}$ is partly compensated by $T_s$ increasing
due to chemical pressure. Also, one may speculate that upon the
approach of the FM QCP, FM fluctuations stimulate triplet SC even
stronger. Obviously, more experiments are needed to unravel the
different pairing and de-pairing contributions to $T_s$.

In summary, magnetic and transport measurements on a series of
polycrystalline UCoGe$_{1-x}$Si$_x$ samples show that
ferromagnetic order and superconductivity are both depressed and
vanish at the same critical concentration $x_{cr} \simeq 0.12$.
The non-Fermi liquid exponent in the resistivity near $x_{cr}$ and
the smooth depression of the ordered moment point to a continuous
FM quantum phase transition. Superconductivity is confined to the
ferromagnetic phase, which provides further evidence for
magnetically mediated superconductivity. These results offer a
unique route to investigate the emergence of superconductivity
near a FM QCP at ambient pressure.

This work was part of the research program of FOM (Dutch
Foundation for Fundamental Research of Matter) and COST Action P16
ECOM.

\end{document}